\documentstyle[twoside,11pt]{article}
\def\greaterthansquiggle{\raise.3ex\hbox{$>$\kern-.75em\lower1ex\hbox{$\sim$}}}
\def\lessthansquiggle{\raise.3ex\hbox{$<$\kern-.75em\lower1ex\hbox{$\sim$}}}
\newcommand{\beq}{\begin{equation}}
\newcommand{\eeq}{\end{equation}}
\newcommand{\beqa}{\begin{eqnarray}}
\newcommand{\eeqa}{\end{eqnarray}}

\newcommand{\ra}{\rightarrow}

\newcommand{\vp}{\varphi}
\newcommand{\vt}{\vartheta}

\def\au{{\setbox0=\hbox{\lower1.36775ex%
\hbox{''}\kern-.05em}\dp0=.36775ex\hskip0pt\box0}}
\def\ao{{}\kern-.10em\hbox{``}}

\normalsize
\begin{document}
\bibliographystyle{plain}


\begin{titlepage}
\begin{flushright}
UWThPh-1992-46\\
September 10, 1992  \\
\end{flushright}
\vspace{3cm}
\begin{center}
{\Large \bf
Saddle point solutions \\[6pt]
in Yang-Mills-dilaton theory}\\[40pt]
Piotr Bizon\footnote{On leave from Institute of Physics, Jagellonian
University, Reymonta 4, 30-059 Cracow, Poland.}\\
Institut f\"ur Theoretische Physik\\
Universit\"at Wien \\
Boltzmanngasse 5, A-1090 Vienna, Austria\\
e-mail: bizon@awirap.bitnet
\vfill
{\bf Abstract}
\end{center}

The coupling of a dilaton to the $SU(2)$-Yang-Mills
field leads to interesting non-perturbative
static spherically symmetric solutions
which are studied by mixed analitical and numerical methods.
In the abelian sector of the theory there are finite-energy
magnetic and electric monopole solutions which saturate the
Bogomol'nyi bound. In the nonabelian sector there exist
 a countable family of globally regular solutions which are purely
magnetic but have zero Yang-Mills magnetic
charge.
 Their discrete spectrum
of energies is bounded from above by the energy of the abelian magnetic
monopole with unit magnetic charge.
 The stability analysis
demonstrates that
the solutions are saddle points of the energy functional with
increasing number of unstable modes.
 The existence and
instability of these solutions are "explained" by the
Morse-theory argument recently proposed by Sudarsky and Wald.
\vfill
\end{titlepage}

\section{Introduction}
As it is well known, the Yang-Mills (YM) equations are scale
invariant which excludes globally regular (i.e. non-singular
with finite energy) static solutions [1,2]. The usual method for
circumventing this nonexistence result is to introduce a Higgs
field. The coupling of the Higgs field has two effects. First,
it breaks scale invariance. Second, a nonabelian gauge group $G$
gets spontanously broken to a subgroup $H$. If the homotopy group
$\pi_{D-1}(G/H)$ is nontrivial (where $D$ is the number of space dimensions),
then the coupled YM-Higgs theory has topologically stable solutions.
A prominent example is the t'Hooft-Polyakov monopole [3] in the $SU(2)$-YM
theory with a triplet Higgs field.

A spontanously broken gauge theory may admit another class of globally
regular solutions if $\pi_{D}(G/H)$ is nontrivial. This homotopy
group is isomorphic to the group of loops in the configuration space
(i.e. space of static, finite energy configurations).
Nontriviality of $\pi_{D}(G/H)$ means that there are
noncontractible loops
passing through the vacuum. The argument, due to Taubes [4] and
Manton [5], of how such noncontractible loops lead to a
nontrivial solution runs as follows. Consider all loops starting
and ending at the vacuum in a fixed homotopy class. On each loop
there is a configuration of maximal energy and the infimum of
these energies gives a saddle point of the energy functional (and
therefore the static solution).
Due to the non-compactness and infinite-dimensinality of the
configuration space, this argument is obviously not rigorous, and
to actually prove that the mini-max procedure converges is a
difficult technical problem. Static solutions corresponding to
saddle points of
the energy functional were called sphalerons to emphasize that,
in contrast to solitons, they are unstable. The existence of a sphaleron
was first shown by Taubes [4] in the $SU(2)$-YM theory with a
triplet Higgs field and by Manton [5] in the $SU(2)$-YM theory
with a complex doublet Higgs field.

Although sphalerons were originally discovered in spontanously
broken gauge theories, it should be stressed that the Higgs
mechanism is by no means necessary for the existence of a sphaleron.
Actually, this is already clear in the $SU(2)$-YM theory with a complex
doublet Higgs, where the gauge group $SU(2)$ is completely
broken and the homotopy group relevant for constructing a
sphaleron is
$\pi_3(SU(2)) \simeq Z$. Thus, in this case the role of the
Higgs field is just to break the scale invariance while the
gauge group itself has a nontrivial third homotopy group. This suggests
that a sphaleron may exist in the $SU(2)$-YM theory coupled to
other fields (of attractive force), provided that the coupling: i) breaks scale
invariance,
and ii) does not alter the topology of the configuration space
of pure $SU(2)$-YM theory.

In this paper I consider a simple example of a coupling which satisfies
these two requirements, namely the coupling of a dilaton. The dilaton,
$\phi$, is a real (massless) scalar field which couples to other
matter fields (with lagrangian $L_m$) through the term $e^{-2a \phi}L_m$,
where $a$ is the dilaton coupling constant. I will show that static
spherically symmetric $SU(2)$-YM-dilaton equations have globally
regular solutions with the following properties:
\begin{enumerate}
\item[a)] there exist a countable family of solutions ${X_n}$
($n\in N$),
\item[b)] the energy $E[X_n]$ increases with $n$ and is bounded
from above,
\item[c)] the solution $X_n$ has exactly $n$ unstable modes.
\end{enumerate}
This family of solutions is in striking analogy to the Bartnik-Mckinnon
(BM) solutions [6] of the Einstein-$SU(2)$-YM equations, which
have the same properties a)-c). In both cases the solution $X_1$
may be interpreted as a sphaleron (for BM solutions this was
first observed by Mazur [7]; see also [8]).

A natural question arises: why do two theories with completely different
dynamics have qualitatively the same spectrum of solutions? The
answer was recently
proposed by Sudarsky and Wald (SW) [9]. They presented a
heuristic but convincing argument which accounts for the properties
a)-c) (except for the boundedness of energy) in the case of BM solutions.
This argument is formulated in the spirit of Morse theory for Hamiltonian
systems and exploits the existence of topologically
nonequivalent multiple
vacua in the $SU(2)$-YM theory (which is related to the fact that
$\pi_3(SU(2)) \simeq Z$. A detailed description of the SW
argument will be
given in Section 7. Here, let me only note that in the case of
the solution $X_1$, the SW argument is, in essence, equivalent
to the mini-max procedure for noncontractible loops. However, in
contrast to the mini-max construction, the SW argument can be
naturally extended (admittedly, under additional assumptions) to
account also for the existence of solutions $X_n$ with $n>1$. Although
the SW argument was originally formulated in the context of
Einstein-YM theory, it is essentially insensitive to the
concrete form of coupling, and applies almost without
modifications to the YM-dilaton theory. In this sense, SW
predicted the
existence of solutions found in this paper. On the
other hand,
the results of this paper give further credence to the SW argument.

The existence of the upper bound for the spectrum may be understood
by considering the $U(1)$ sector of the YM-dilaton theory.
Surprisingly enough, there are {\it finite} energy abelian
solutions which describe magnetic and electric point monopoles.
The finiteness of energy is due to the regulating effect of the dilaton
which weakens the short distance singularity. Morever, these solutions
saturate the Bogomol'nyi bound (in the $U(1)$ sector), hence
their energies are equal to their charge. It turns out from
numerics that the limiting solution $X_\infty$ (whose energy
bounds the
spectrum from above) corresponds to the abelian magnetic
monopole with unit magnetic charge.

The YM-dilaton theory and the Einstein-YM theory may
be embedded in a single Einstein-YM-dilaton theory governed by the action
\beq
S = \int d^4 x \sqrt{-g} \left[-\frac{1}{G} R +2 (\nabla \phi)^2
+ e^{-2a \phi} F^2 \right] \;.
\eeq
This theory is characterized by a dimensionless parameter
$\alpha=a/\sqrt{G}$.
When $\alpha \ra \infty$, the action (1) reduces to the YM-dilaton
theory. When $\alpha=0$, the action (1) becomes the
Einstein-YM theory (plus trivial kinetic term for the scalar field).
 Finally, the case $\alpha=1$ corresponds to
the low-energy string theory. It was shown by the author
elsewhere [10] that the theory defined by the action (1) has
static spherically symmetric (globally regular and black-hole) solutions
with properties a)-c), for all values of $\alpha$. This paper
specializes to the limiting case $\alpha \ra \infty$. It seems
instructive to consider this case separately, because it involves
the essential features of the general case, but has an
advantage of being considerably simpler, which allows to obtain
some analitical estimates on the parameters of solutions. Also,
the non-perturbative effect of the dilaton can be clearly seen in
this model.

The paper is organized as follows. In the next Section the field
equations are derived and some scaling properties are discussed.
In Section 3 the explicit abelian solutions are described.
In Section 4 the a priori behaviour of globally regular
solutions is obtained.
 In Section 5 the numerical results are presented and
some qualitative properties of solutions are discussed. The deep
analogy between these solutions
and the BM solutions is emphasized. Section
6 is devoted to the stability analysis. Finally, in Section 7 the SW
argument is summarized and some possibilities
of proving rigorously the existence of numerical solutions are suggested.

\section{Field equations}
The dynamics of the $SU(2)$-YM field coupled to a dilaton is
defined by the action
\beq
S = \int d^4 x \left[ 2 (\nabla \phi)^2 + e^{-2a \phi} F^2 \right] \;,
\eeq
where $F = dA + e A \wedge A$ is the YM curvature of the $SU(2)$
connection $A$ and $\phi$ is the dilaton. Hereafter, for
convenience I put the coupling constants $a=e=1$, which is equivalent
to choosing $a/e$ as the unit of length and $1/ae$ as the unit
of energy.

The field equations derived from (2) are
\beq
D ( e^{-2 \phi} \ast \! F ) = 0 \;,
\eeq
\beq
{\nabla}^2 \phi + \frac{1}{2} e^{-2 \phi} F^2 = 0 \;,
\eeq
where $D$ is the $SU(2)$ covariant derivative.

I wish to find static spherically symmetric solutions to these equations
that are globally regular i.e. non-singular and with finite energy.

The most general spherically symmetric $SU(2)$  connection has the form
[11]
\beq
A = a \tau_3 dt + b \tau_3 dr + (w \tau_1 + d\tau_2) d\vt +
(\cot \vt \tau_3 + w \tau_2 - d\tau_1) \sin \vt d\vp\;\; ,
\eeq
where $a$, $b$, $w$ and $d$ are functions of $(r,t)$ and $\tau_i$
$(i = 1,2,3)$ are generators of $su(2)$ Lie algebra. Using the residual
gauge freedom the radial gauge $b = 0$ can be imposed. When the connection
is static, i.e.
$a$, $w$ and $d$ depend only on $r$, one can also set $d = 0$ by a constant
gauge transformation. Hence, the general static spherically symmetric
$SU(2)$ connection is described by two functions: the electric
potential $a(r)$ and the magnetic potential $w(r)$. Now, I
assume further that $a \equiv 0$. Actually, this is not a
restriction because one can show, following the argument given
in [12], that there are no globally regular solutions with
nonzero electric field.

The purely magnetic YM curvature is
\beq
F =  w' \tau_1 dr \wedge d\vt + w' \tau_2 dr \wedge \sin \vt d\vp
 - (1 - w^2) \tau_3 d\vt \wedge \sin \vt d\vp \;,
\eeq
where prime denotes differentiation with respect to $r$. For $F$
given by this ansatz and for $\phi = \phi(r)$, the equations (3)
and (4) reduce to the following system
\beq
(e^{-2\phi} w')' +  \frac{1}{r^2} e^{-2 \phi} w(1 - w^2) = 0\; ,
\eeq
\beq
(r^2 \phi')' + 2 e^{-2\phi} \left[ w'{}^2 + \frac{(1-w^2)^2}{2r^2} \right] = 0.
\eeq
These equations may also be derived from the variation of the energy
functional
\beq
E[w,\phi] =  4 \pi \int_{0}^\infty T_{00}\; r^2 dr\;\;,
\eeq
where $T_{00}$ is the local energy density
\beq
4 \pi T_{00} = \frac{1}{2} \phi'{}^2 +
e^{-2\phi} \left[  \frac{1}{r^2} w'{}^2 + \frac{(1 - w^2)^2}{2r^4} \right] \;.
\eeq
Let me make two remarks which will be useful in the subsequent discussion.
First, note that, in general, the energy functional $E$ is extremized
only against variations with $\delta \phi(\infty)=0$. However,
it is also useful to  consider more general variations for which
$\delta \phi(\infty) \neq 0$. Then,
the variation of energy around a solution has the form
\beq
\delta E = D \; \delta \phi(\infty) \;,
\eeq
where $D = \lim_{r \ra \infty} r^2 \phi{}'$ is the dilaton charge.
To avoid confusion I want to emphasize that the "surface term"
on the right side of eq.(11) is not of the Regge-Teitelboim type
(in particular it cannot be cancelled by adding correction to energy)
but it is rather a term which appears in variational problems
with a free end.

Second, note that the equations (7) and (8) have a "scaling" symmetry.
Namely, if $w(r)$ and $\phi(r)$ are solutions so are
\beq
\begin{array}{l}
w_{\lambda}(r) = w(e^{\lambda} r) \;,\\[12pt]
\phi_{\lambda}(r) = \phi(e^{\lambda} r) + \lambda \;.
\end{array}
\eeq
Under this transformation the energy scales
as follows
\beq
E[w_{\lambda},f_{\lambda}] = e^{-\lambda} E[w,f] \;.
\eeq
The existence of this "scaling" symmetry excludes, via Derrick's
argument, nontrivial static finite energy solutions with vanishing
dilaton charge $D$ (nota bene this also follows immediately from
eq.(8)). However, when $D \neq 0$, Derrick's argument doesn't
apply because for the variation induced by the transformation (12)
$\delta \phi(\infty)$ is nonzero, and therefore, as follows from
(11), the energy is not extremized. Hereafter, I will assume
that all solutions satisfy $\phi(\infty)=0$, which can always be
set by the transformation (12). This choice sets the scale of
energy in the theory.

\section{Abelian solutions}
The equations (7) and (8) have two explicit abelian solutions. The first
one is the vacuum solution
\beq
w = \pm 1 \;\;, \qquad \qquad \phi = 0
\eeq
for which the energy has the global minimum $E=0$.

The second solution is
\beq
w = 0 \;\;, \qquad \qquad \phi = \ln (1 + \frac{1}{r})
\eeq
and its YM curvature is
\beq
F = - \tau_3 \; d\vt \wedge \sin \vt d\vp \;,
\eeq
which corresponds to the Dirac magnetic monopole with unit
magnetic charge. There is also an
electrically charged abelian solution related to (15)
 by the duality rotation :
\beq
\tilde{F} = e^{-2 \phi} \ast \! F = \frac{1}{(1+r)^2} \tau_3
\; dr \wedge dt \;, \quad \quad  \tilde{\phi} = - \phi \;.
\eeq
These solutions
have very interesting properties.
The dilaton dramatically changes the properties of $U(1)$ point monopoles
(electric and magnetic). Without a dilaton, the energy density
of a point monopole diverges at $r=0$ as $T_{00} \sim 1/r^4$,
whereas in the present case $T_{00} \sim 1/r^2$. Thus, although
the solution (15) is singular at $r=0$, its total energy is finite
and equals one!
This result may be viewed as non-perturbative cancellation of two
infinite self-energies: positive one of the point magnetic monopole
and negative one of the dilaton.

 Since this solution will play an important role
in the discussion of nonabelian solutions, it is useful to see
how one can obtain it in a systematic way. Namely, in the $w
\equiv 0$ sector, the energy (9) can be written as
\beq
E[\phi] =  \int_{0}^\infty  ( r\phi{}' + \frac{1}{r} e^{-\phi} )^2\; dr\;+
e^{-\phi}\mid_{0}^\infty \;.
\eeq
Thus, if $\phi(0)=\infty$ (and $\phi(\infty)=0$), the energy is bounded
from below by the value of magnetic charge (here set equal to one)
and attains a global minimum $E=1$
on solutions of the first order Bogomol'nyi-type equation
\beq
r^2 \phi{}' + e^{-\phi} = 0 \;.
\eeq
Solutions of this equation automatically satisfy the eq.(8) with
$w=0$. Elementary integration of this equation gives the
solution (15). A detailed discussion of Bogomol'nyi inequalities
in the Maxwell-dilaton theory, without an assumtion of spherical
symmetry, will be given elsewhere [13].

\section{Boundary conditions}
Now
I will specify the boundary conditions for the globally
regular solutions. They are determined by the requirement that
\begin{enumerate}
\item[i)] the local energy density be finite for all $r$
\beq
T_{00} < const < \infty
\eeq
\end{enumerate}
which imposes boundary conditions for $w$ and $\phi$ at $r=0$, and
\begin{enumerate}
\item[ii)] the energy be finite
\beq
E < \infty
\eeq
\end{enumerate}
which imposes boundary conditions for $w$ and $\phi$ at infinity.

It is easy to construct asymptotic solutions to eqs. (7) and (8)
satisfying these boundary conditions. The solution near $r=0$ is
\beq
\begin{array}{l}
w = 1 - br^2 + O(r^4) \;,\\[12pt]
\phi = c - 2b^2r^2 + O(r^4) \;.
\end {array}
\eeq
At $r = \infty$ the asymptotic solution is
\beq
\begin{array}{l}
\pm w = 1 - d/r + O(1/r^2) \;,\\[12pt]
\phi = e/r + O(1/r^4) \;.
\end{array}
\eeq
Here $b,c,d$, and $e$ are arbitrary constants.
All higher order terms in the above
expansions are uniquely determined, through recurence relations,
by $b$ and $c$ in (22), and $d$ and $e$ in (23).

Using these boundary conditions one can get the following elementary
a priori results for the solutions of eqs.(7) and (8):
\paragraph{} {\em Lemma 1.} The function $w$ oscillates around zero
between $-1$ and $1$ (or $|w|\equiv 1$).

{\em Proof.} It follows from eq.(7) that if $w'(r_0)=0$ then
at $r_0$
\beq
sgn \; w'' = sgn \; w(w^2-1) \;.
\eeq
This implies that $w$ cannot have local maxima for $w>1$ and
local minima for $w<-1$. Since $w(0)=1$ and $|w(\infty)|=1$,
this gives $|w|<1$ for all $r>0$. Thus from (24),
$w''w<0$, which concludes the proof.
\paragraph{} {\em Lemma 2.} The function $\phi$ is monotonically decreasing.

{\em Proof.} As above, this follows immediately from the
maximum principle applied to eq.(8).
\paragraph{}

Finally, note that, for the asymptotic behaviour (23), the
radial magnetic curvature, $B = \tau_3 (1-w^2)/r^2$, falls-off
as $1/r^3$, and therefore all globally regular solutions have
zero YM magnetic charge.

\section{Nonabelian solutions}
Let us assume that there exist a 2-parameter family of local
solutions defined by the expansion (22). Note that this is a
nontrivial statement because the point $r=0$ is a singular point
of the equations (7) and (8), so the formal power-series
expansion (22) may have, in principle, a zero radius of convergence.
A generic solution with initial data (22) certainly will not
satisfy the asymptotic conditions (23) (in fact, the solution
may even become singular at some finite distance). The standard numerical
startegy, called the shooting method, is to find
initial data $(b,c)$ for which the local solution extends to a
global solution with the asymptotic behaviour (23). Actually, only
$b$ is a nontrivial shooting parameter, since one take arbitrary
$c$ and after finding the solution adjust the value of $\phi(\infty)$
to zero using the transformation (12). For generic orbits with
$b<b_{\infty} \simeq 0.3795$ the
function $w$ oscillates finite number of times
in the region between $w=-1$ and $w=1$ and then goes to $\pm \infty$.
For $b>b_{\infty}$
all orbits become singular at a
finite distance (in a sense that $w'$ becomes infinite).

The numerical results strongly indicate that there exist a countable
family of initial data $(b_n,c_n)$, $ n \in N$, determining globally
regular
solutions $X_n=(w_n,\phi_n)$. Here the index $n$ labels the number
of nodes of the function $w$.
The values of initial data and
energies of the first five
solutions are given in Table 1.
The functions $w$ and $\phi$ for the first three solutions are graphed
in Fig.1 and 2.

\begin{table} [h]
\caption{Initial data $(b,c)$ and energies of the first five globally
regular solutions.}
$$
\begin{tabular}{|c|c|c|c|} \hline
$n$ & $b$ & $c$ & $E$ \\ \hline
1 & 0.26083011 & 1.711 & 0.804 \\
2 & 0.35351804 & 3.374 & 0.9659 \\
3 & 0.3750017 & 5.158 & 0.9944 \\
4 & 0.378754 & 6.966 & 0.9992 \\
5 & 0.379373 & 8.754 & 0.99993 \\  \hline
\end{tabular}
$$
\end{table}

 The solutions display three characteristic regions. The energy density
$T_{00}$ is concentrated in the inner core region $r<R_1$, where $R_1$ is
approximately the location of the first zero of $w$. This region
decreases with $n$ and shrinks to zero as $n \ra \infty$. In
the second region, $R_1 < r <R_2$, where $R_2$ is approximately the
location of the last but one zero of $w$, the function $w$ slowly
oscillates around $w=0$ with a very small amplitude. In this region
the solution is very well approximated by the abelian magnetic
monopole (15). This region extends
to infinity as $n \ra \infty$. Finally, in the asymptotic region
 $r>R_2$, the function $w$ goes
monotonically to $w=\pm 1$ (hence the YM magnetic charge is gradually screened)
and for $r \ra \infty$ the solution tends to the vacuum ($w=\pm
1$, $\phi=0$).

Because of these properties, the solutions are in striking
resemblance to the BM solutions [6] of the Einsten-YM equations
 - the dilaton coupling has
almost the same effect as the gravitational coupling. In both cases
the equilibrium configurations result from a balance between repulsive
YM force and attractive, gravitational or dilatonic, force.
There are indications that this analogy is even deeper. Below I will
discuss two facets of the apparent duality between gravity and dilaton
interacting with the YM field.

First, I will show that in the YM-dilaton theory the
energy of a static solution can be expressed as a surface integral at spatial
infinity.
To show this, I will first derive a simple virial identity. Consider
a one-parameter family of field configurations defined by
\beq
\begin{array}{l}
w_{\beta}(r) = w(\beta r)\;,\\[12pt]
\phi_{\beta}(r) = \phi(\beta r)\;.
\end{array}
\eeq
For this family the energy (9) is given by
\beq
E[w_{\beta},\phi_{\beta}] = \beta^{-1} I_1 + \beta\; I_2
\eeq
where
\beq
I_1 = \frac{1}{2} \int_{0}^\infty r^2  \phi{}'{}^2 dr \;,
\eeq
\beq
I_2 = \int_{0}^\infty e^{-2\phi} \left[ w'{}^2 + \frac{(1 -
w^2)^2}{2r^2} \right] dr \;.
\eeq
Since the energy is extremized at $\beta=1$, it follows from
(26) that on-shell
\beq
I_1 = I_2 \;.
\eeq
Next, integrating eq.(8) one gets $D=-2 I_2$ ($D$ is the dilaton
charge defined in Section 2), and therefore
eq.(29) yields
\beq
E = -D\;.
\eeq
Thus the energy of a static
solution can be read-off from the monopole term of the
asymptotic expansion (23) of the dilaton field. This is a remarkable
property which reminds very much the situation in general
relativity and shows a relation between the dilaton field
and the conformal degree of freedom of the metric.

Secondly, the most striking analogy between our solutions and the
BM solutions is their spectrum of energies (see Table 1 and compare
with Table I in ref.[14]). In both cases the energies increase with
$n$ and are bounded from above by $E=1$. This cannot be a coincidence,
but what distinguishes this particular value of energy which provides
the common upper bound? The answer is remarkably simple. The limiting
$X_{\infty}$ solutions (whose energies give upper bounds) of our
family and the BM family saturate the Bogomol'nyi inequalities
in the abelian sectors of respective theories and therefore their energies are
equal to
the unit magnetic charge. To see this, consider first the dilatonic
solutions. As was discussed above, the second region $R_1<r<R_2$,
covers the whole space as $n \rightarrow \infty$, since
in this limit $R_1 \rightarrow 0$ and
$R_2 \rightarrow \infty$. As $n$ grows the amplitude of oscillations of the
function $w$ decreases and goes to zero as $n \rightarrow \infty$.
Thus, for $n \rightarrow \infty$ the solution $X_n$ tends (nonuniformly)
to the (singular) abelian magnetic monopole described in Section
3:
\beq
X_{\infty} = (\; w = 0, \; \phi = \ln (1+\frac {1}{r})\; ) \;.
\eeq
As I have shown in Section 3, in the $U(1)$ sector of the YM-dilaton theory,
the
static solutions satisfy the Bogomol'nyi inequality $E \geq Q$.
The limiting solution $X_{\infty}$ saturates the bound in the Q=1
sector.

The behaviour of the BM solutions (in isotropic coordinates)
is similar: as $n \rightarrow \infty$ the YM field tends to the abelian
magnetic monopole
while the metric develops a horizon and becomes the extremal Reissner-Nordstrom
black hole solution with unit magnetic charge. Thus
\beq
X_{\infty}^{BM} = \left( w=0, \;ds^2 = -e^{-2 U} dt^2 + e^{2 U} (dx^2 + dy^2 +
dz^2)\right) \;,
\eeq
where
\beq
U = \ln (1 + \frac{1}{r}) \;.
\eeq
It is well known that this solution saturates the Bogomol'nyi inequality in
 Einstein-Maxwell theory [15]. Actually, the limiting solutions
(31) and (32) can be mapped one into another by the duality transformation
$U \leftrightarrow \phi$ and $\alpha \leftrightarrow 1/{\alpha}$
in the abelian sector of the theory defined by the action (1).

{}From the content of the last two paragraphs it is clear that to prove
rigorously that the energy is bounded from above by one, one needs
in the YM-dilaton theory (and in the Einstein-Yang-Mills theory in the case
of BM solutions)
a sort of Bogomol'nyi inequality with {\em reversed} sign, $E\leq Q$, which
is saturated by the limiting abelian solution. Unfortunately, I
wasn't able to find
 such an inequality. It would be probably easier,
 but also much less interesting, to find an upper bound which is
not sharp (for the BM solutions that was done in ref.[14]).
Also, it
is not difficult to obtain not strict bounds on initial parameters.
For example, multiplying eq.(8) by $\phi$, integrating by parts
and combining the result with eq.(29) yields the identity
\beq
\int_{0}^\infty (\phi - 1) e^{-2\phi} \left[ w'{}^2 + \frac{(1 -
w^2)^2}{2r^2} \right] dr = 0 \;,
\eeq
which implies that $\phi(0)\geq 1$.

\section{Stability analysis}
In this Section I address the issue of linear stability of the
static solutions described above. To that purpose one has to
analyse the time evolution of linear perturbations about the equilibrium
configuration. I will assume that the time-dependent solutions
remain spherically symmetric and the YM field stays within the
ansatz (6). This is sufficient to demonstrate instability
because unstable modes appear already in this class of perturbations.
The spherically symmetric evolution equations are
\beq
-(e^{-2 \phi} \dot{w})\dot{} + (e^{-2 \phi} w')' + \frac{1}{r^2} e^{-2 \phi}
w(1 - w^2) = 0 \;,
\eeq
\beq
-r^2 \ddot{\phi} + (r^2 \phi{}')' + 2 e^{-2 \phi} \left[ {w'}^2 +
\frac{(1-w^2)^2}{2r^2} \right] = 0 \;,
\eeq
where dot denotes differentiation with respect to time $t$.

Now, I take the perturbed fields: $w(r)+\delta w(r,t)$, and
$\phi (r)+\delta \phi (r,t)$, where $(w(r),f(r))$ is a static solution,
and insert them into the eqs. (35),(36). Linearizing and
assuming the harmonic time-dependence for the perturbations:
$\delta w(r,t)=e^{i\sigma t} \xi (r)$ and
$\delta \phi (r,t)=e^{i\sigma t} \psi (r)$, one obtains an
eigenvalue problem
\beq
-\xi{}'' + 2 \phi{}' \xi{}' + 2 w' \psi{}' - \frac{1}{r^2} (1-3 w^2) \xi
= {\sigma}^2 \xi \;,
\eeq
\beq
-(r^2 \psi{}')' - 4 e^{-2\phi} \left[ w'\xi{}' - \frac{1}{r^2}
w(1-w^2) \xi - ( {w'}^2 + \frac{(1-w^2)^2}{2r^2} ) \psi \right] =
{\sigma}^2 r^2 \psi \;.
\eeq
It is easy to check that if the perturbations satisfy the
boundary conditions
\beq
\begin{array}{l}
\xi (0) = 0 \; \qquad \qquad \psi (0) = const \;,\\[12pt]
\xi (\infty) = 0 \; \qquad \qquad \psi (\infty) = 0 \;,
\end{array}
\eeq
then the above system is self-adjoint, hence eigenvalues
${\sigma}^2$ are real. Instability manifests itself in the
presence of at least one negative eigenvalue.

To solve the eigenvalue equations (37),(38) with the boundary conditions
(39), is a straighforward but tedious numerical problem. I have
done that
for several lowest-energy static solutions. It turns out that
the solution $X_1$ has exactly one unstable mode of frequency
 ${\sigma}^2 \simeq -0.0225$. Each succesive static solution
picks up one additional unstable mode (I have checked this up to $n=4$).
This is consistent with the fact that the limiting solution $X_{\infty}$,
given by (31), has infinitely many unstable modes. To prove this, consider the
perturbations of $X_{\infty}$ with $\delta \phi = 0$. Then,
eq.(37) reads
\beq
-\xi{}'' - \frac{2}{r(1+r)} \xi{}' - \frac{1}{r^2} \xi = \sigma^2
\xi \;.
\eeq
This equation has infinitely many negative modes because the zero
energy solution satisfying $\xi(0)=0$, has infinitely many nodes
as can be seen easily from the asymptotic solution.

The result that
 the solution $X_n$ has $n$ unstable modes
 fits very nicely to the interpretation
of solutions. In particular, for the interpretation of the
solution $X_1$ as a sphaleron, it is crucial that it has exactly
one unstable mode. However, remember that I have considered a restricted
class of perturbations, and by doing so some directions of instability
might have been supressed. If there are additional unstable
modes outside the ansatz (which I doubt), the interpretation of
solutions given by Sudarsky and Wald would have to be revised.

\section{Sudarsky and Wald's argument}
Sudarsky and Wald have recently proposed a heuristic argument
which "explains" the existence and instability of the BM
solutions of the Einstein-YM equations. This argument is, in principle,
applicable to other theories involving the $SU(2)$-YM field,
which are not scale invariant and possess a stable solution.
Below I outline the SW argument in application to the $SU(2)$-YM-dilaton
theory.

Let $\tilde{\Gamma}$ be a space of all functions $(A_i,\phi)$, defined
over $R^3$, for which the energy $E$ is finite. Let $\overline{\Gamma}$ be
a subspace of $\tilde{\Gamma}$ with $\phi(\infty)=0$. The static
solutions are extrema of energy on $\overline{\Gamma}$. One such extremum
is the vacuum solution $(A_i,\phi=0)$ for which energy has a
global minimum $E=0$. Now, the key feature of the $SU(2)$-YM
group is the presence of "large gauge transformations" i.e.
topologically inequivalent cross-sections of the YM-bundle,
classified by the homotopy group $\pi_{3}(SU(2)) \simeq Z$.
Thus, the energy functional $E$ has a countable set of
disconnected global minima corresponding to the trivial vacuum
$(A_i,\phi=0)$ and all large gauge transformations of it. To
avoid complications with the group of small gauge
transformations $G$, it is convenient to pass from $\overline{\Gamma}$ to
the space of gauge orbits $\Gamma=\overline{\Gamma}/G$.

Now, to apply Morse theory methods in Banach spaces, one needs a
sort of compactness condition (like the Palais-Smale condition).
A convenient way of implementing such a condition (which is here
simply assumed to hold), is to introduce on $\Gamma$ a
Riemannian metric $G_{AB}$ (upper case latin letters denote
indices of tensor fields on $\Gamma$), such that the flow
generated by the vector field $M^{A}=-G^{AB}{\nabla}_{B}E$
carries each point of $\Gamma$ to a critical point of $E$. As discussed
above, there exist a countable set of global minima of $E$.
Since this set is disconnected, the flow defined by $M^{A}$
cannot carry all points of $\Gamma$ to global minima (or local
minima if any exist), because this would contradict the
connectedness of $\Gamma$. Thus, the set, $\Gamma_1 \subset
\Gamma$, of points
which do not flow to local minima, must contain at least one
critical point of $E$. A critical point on $\Gamma_1$ with
smallest energy $E_1$ is a saddle point on $\Gamma$ with exactly
one unstable direction. This is believed to account for the
existence of the solution $X_1$.

Actually, there is a countable set of local minima of $E$
restricted to $\Gamma_1$, namely $X_1$ and all large gauge transformations
of it. Hence, one can repeat the above argument, replacing
$\Gamma$ by $\Gamma_1$ (and assuming that $\Gamma_1$ is
connected), to predict the existence of a submanifold $\Gamma_2
\subset \Gamma_1$ with a point $X_2$, whose energy $E_2$
minimizes $E$ restricted to $\Gamma_2$. The point $X_2$ is an
extremum of $E$ on $\Gamma$, which has the 2-dimensional space
of unstable directions. This is believed to account for the
existence of the second static solution $X_2$. All higher $n$
solutions are predicted by the repetition of this argument.

It seems very unlikely that the SW argument in its
present form can be converted into a rigorous proof. However, the
same argument can be made for spherically symmetric connections.
Then, the powerful methods of equivariant Morse theory are
available, and
in fact these methods were succesfully applied in related
problems [16]. In my opinion this is
a very promising direction for future research.

Another possibility of proving rigorously the existence of numerical
solutions found in this paper
 is to apply the
methods of the dynamical systems theory. This approach was used
recently by Smoller and his collaborators [17,18], who proved the
existence of the BM family of solutions to the Einstein-YM equations.
 A similar
proof should be possible for the YM-dilaton equations although
it might be more difficult, because here the corresponding (nonautonomous)
dynamical system is four-dimensional whereas in the Einstein-YM
case it is three-dimensional.

\paragraph{Acknowledgement.}
I am grateful to Peter Aichelburg for his continuous interest in
my work and many discussions. I thank Robert Beig, Gary Horowitz,
Max Meinhart, Walter Simon
and Robert Wald for useful comments.
I would like to acknowledge the hospitality of the Aspen Center for Physics,
 where part of this work was carried out.
This work was supported by the Fundaci\'on Federico.

\section*{Figure captions}
\begin{description}
\item[Fig.1] The function $w$ for the first three globally regular
solutions.
\item[Fig.2] The function $\phi$ for the first three globally regular
solutions.
\end{description}

\newpage


\newpage
\begin{thebibliography}{99}
\bibitem{1} S.Coleman, in New Phenomena in Subnuclear Physics,
ed. A.Zichichi (Plenum, New York, 1975).
\bibitem{2} S.Deser, Phys.Lett. {\em 64 B} (1976) 463.
\bibitem{3} G.t'Hooft, Nucl.Phys. {\em B279} (1974) 276;
A.M.Polyakov, JETP Lett. {\em 20} (1974) 194.
\bibitem{4} G.H.Taubes, Comm.Math.Phys. {\em 86} (1982) 257; 299;
ibid {\em 97} (1985) 473.
\bibitem{5} N.S.Manton, Phys.Rev. {\em D28} (1983) 2019.
\bibitem{6} R. Bartnik and J. Mckinnon, Phys, Rev. Lett. {\em 61} (1988) 41.
\bibitem{7} P.O.Mazur, private communication, 1990.
\bibitem{8} D.V.Gal'tsov and M.S.Volkov, Phys.Lett. {\em B273}
(1991) 255.
\bibitem{9} D.Sudarsky and R.M.Wald, Phys.Rev. {\em D46} (1992) 1453.
\bibitem{10} P.Bizon, {\em Saddle point solutions in low energy string
theory}, to appear
\bibitem{11} E.Witten, Phys.Rev.Lett. {\em 38} (1977) 121.
\bibitem{12} P.Bizon and O.T.Popp, Class.Quantum Grav. {\em 9}
(1992) 193.
\bibitem{13} P.Bizon, {\em Bogomol'nyi inequalities in Maxwell-dilaton
theory}, to appear
\bibitem{14} H.P.K\"unzle and A.K.M. Masood-ul-Alam,
J.Math.Phys. {\em 31} (1990) 928.
\bibitem{15} G.W.Gibbons and C.M.Hull, Phys.Lett. {\em 109B}
(1982) 190.
\bibitem{16} L.M.Sibner et al., Proc.Natl.Acad.Sci. {\em 86}
(1989) 8610.
\bibitem{17} J.Smoller et al., Comm.Math.Phys. {\em 143} (1991) 115.
\bibitem{18} J.A.Smoller and A.G.Wasserman, University of Michigan preprint,
 1992.
\end{thebibliography}
\end{document}